\documentstyle[twoside,fleqn,espcrc2,epsfig]{article}

\newcommand{\mtx}[1]{{\mathsf #1}}

\begin{document}

\thispagestyle{empty}

\title{Numerical Stability of Lanczos Methods \hfill %
       \raisebox{10.0 ex}[1.0 ex][0.0 ex]{%
         \parbox[t]{1.5in}{
           \flushright \small
           LTH457\\
           TCDMATH 99-10\\
           hep-lat/9909131}}}

\author{Eamonn Cahill,
        \address{
        School of Mathematics, Trinity College, Dublin 2, Ireland
        and 
        Hitachi Dublin Laboratory, Dublin 2, Ireland}
        Alan Irving,         
        \address{
        Theoretical Physics Division, Department of Mathematical Sciences,
        University of Liverpool, PO Box 147, Liverpool L69 3BX, UK}
        Christopher Johnson, \hbox{$^{\rm b}$}
        James Sexton, \hbox{$^{\rm a}$} 
        and The UKQCD Collaboration}

\begin{abstract}
The Lanczos algorithm for matrix tridiagonalisation suffers from
strong numerical instability in finite precision arithmetic when
applied to evaluate matrix eigenvalues.  The mechanism by which this
instability arises is well documented in the literature. A recent
application of the Lanczos algorithm proposed by Bai, Fahey and Golub
allows quadrature evaluation of inner products of the form
$\psi^\dagger\cdot g(\mtx{A})\cdot\psi$.  We show that this quadrature
evaluation is numerically stable (accurate to machine precision) and
explain how the numerical errors which are such a fundamental element
of the finite precision Lanczos tridiagonalisation procedure are
automatically and exactly compensated in the Bai, Fahey and Golub
algorithm.  In the process, we shed new light on the mechanism by
which roundoff error corrupts the Lanczos procedure.
\end{abstract}

\maketitle


\section{THE BAI, FAHEY AND GOLUB METHOD}

The determinant of the fermion matrix plays a central role in
dynamical Lattice QCD simulations.  The direct
approach \cite{ref:duncan}
to evaluating
fermion determinants typically makes use of the Lanczos procedure
for eigenvalue estimation of Hermitian matrices.  Given an $n\times n$
Hermitian matrix $\mtx{A}$, and an orthonormal seed vector $f_1$, the
Lanczos procedure is an iterative process which (in exact arithmetic)
generates a sequence of orthogonal vectors $f_i, i=1,\ldots,m$.  If we
define an $n\times m$ matrix $\mtx{F}_m = (f_1, \ldots, f_m)$ whose
columns are given by the vectors $f_i$, then we have:
$\mtx{F}_m^\dagger \mtx{F}_m = \mtx{I}_m$, 
$\mtx{T}_m = \mtx{F}_m^\dagger \mtx{A} \mtx{F}_m$ is
tridiagonal,
and the eigenvalues of 
$\mtx{T}_m$ converge to the eigenvalues of $\mtx{A}$ as $m\rightarrow n$.  

An alternative
stochastic approach to evaluating determinants
was proposed in 1996 by Bai, Fahey and Golub (BFG) 
\cite{ref:bfg} who observe that
a vector-matrix-vector of the form $\psi^\dagger \cdot g(\mtx{A}) \cdot \psi$ 
can be expressed as an integral of the function $g(\cdot)$ over a modified
spectral measure. 
An $m$-point
Gaussian quadrature approximation to this integral is then given by
$\psi^\dagger \cdot g(\mtx{A}) \cdot \psi \approx
\sum_{i=1}^m w_i g(\theta_i).$
The remarkable result \cite{ref:davis} applied in the BFG method
is that the abscissa, $\theta_i$,
of this quadrature rule are
given by the eigenvalues of the tridiagonal matrix $\mtx{T}_m$ which
is generated by a Lanczos procedure applied with seed vector $\psi$, 
and that 
the weights, $w_i$, are given as the squares of the first components of the
corresponding eigenvectors of $\mtx{T}_m$.

\begin{figure}[htb]
        \centering
        \epsfig{file=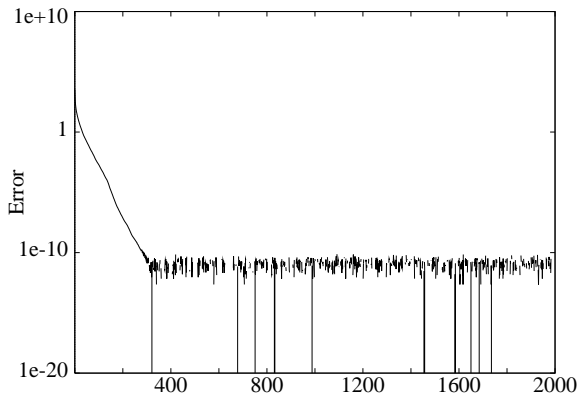,width=7.5cm}
        \caption{Convergence history for the evaluation of 
                 $\psi^\dagger \cdot \log(\mtx{M}_\kappa^\dagger \mtx{M}_\kappa) \cdot \psi $ 
                 on a $12^3\times 24$ quenched configuration as a function of the number
                 of Lanczos iterations.}
        \label{fig:error2000}
\end{figure}

Figure~\ref{fig:error2000} shows an example of the results which the
BFG method generates on a $12^3\times 24$ quenched gauge
configuration.  The configuration is a quenched configuration
with $\beta=5.7$ produced within the current UKQCD 
research program.  A random Gaussian fermion vector, $\psi$, was
generated, and the BFG method applied to evaluate
$\psi^\dagger\cdot\log(\mtx{M}_\kappa^\dagger\mtx{M}_\kappa)\cdot\psi$
for Wilson hopping parameter $\kappa = .1650$ as a function of the
order $m$ of the Lanczos tridiagonal matrix and of the resulting
Gaussian quadrature approximation.  To generate an estimate of the
``true'' value for this matrix-vector inner-product for comparison, we
arbitrarily averaged the values for order 1991 to order 2000.  The
plot then shows the log of the difference of the $m$-th order BFG
estimate and this ``true'' value plotted against $m$.  The plot shows
a rapid convergence to the ``true'' value (convergence has occurred at
about order 200), followed by what is clearly fluctuations at the
level of algorithm precision.  Note in particular that the long
converged region is perfectly flat, and shows no drift.  Checks for a
toy model, and the mathematical analysis which we briefly describe
here, indicate that the convergence is in fact, convergence to the
exact answer, and that finite precision arithmetic introduces no
systematic error.

\section{THE EFFECT OF FINITE PRECISION ARITHMETIC}

At first sight, Figure~\ref{fig:error2000} might be interpreted as
good phenomenological evidence that the BFG method is reliable and
robust.  A second sight however should disturb a person familiar with
the Lanczos procedure at finite precision.  In finite precision
arithmetic, the Lanczos procedure generates clusters of almost
degenerate eigenvalues of the tridiagonal matrix $\mtx{T}_m$ where
only a single eigenvalue would have occurred in an infinite precision
calculation.  These clusters could potentially destroy the convergence
of the BFG quadrature sum to the correct infinite precision result
since the corresponding abscissa are included multiple times rather
than singly in that sum.  We have found however, that when clusters
arise, the corresponding weights in the quadrature rule adjust so that
the net contribution of the cluster of multiple eigenvalues and weights is
equal to that of the corresponding single eigenvalue and
weight which would arise in infinite precision.

\begin{table*}[ht]
\centering
\caption{This table illustrates the invariance of the sum of the weights for an
eigenvalue cluster generated from a finite precision Lanczos procedure.}
\medskip
\label{tab:clust_wts}
\begin{tabular}{llll}
\hline 
$m$  & Weights in Cluster     & Sum of Weights & Relative Error    \\
\hline 
40 
&  0.121274864637753313E-02 & 0.121274864637753313E-02 & 0.250E-14 \\
120 							      
&  0.239992265988143331E-04 & 0.121274864637753270E-02 & 0.286E-14 \\
&  0.298384753074422057E-03 & \\			      
&  0.890364666704296305E-03 & \\			      
160 							      
&  0.128211917795651532E-05 & 0.121274864637753248E-02 & 0.303E-14 \\
&  0.118900026992380976E-03 & & \\			      
&  0.470001178219875870E-03 & & \\			      
&  0.622565321987319131E-03 & & \\			      
\hline
\end{tabular}
\end{table*} 

This effect is illustrated in Table~\ref{tab:clust_wts}
which shows the weights
corresponding to a given cluster for a Lanczos procedure applied to a model
problem which has been chosen so all quantities can be calculated both 
numerically and exactly and direct comparisons can be made.
At iteration 40, the cluster in question contains only a single eigenvalue,
and the relative error between the value of the weight as generated by the Lanczos procedure
and an exact calculation is seen to be $O(10^{-14})$.
By iteration 160, the cluster expands to include four almost degenerate eigenvalues
(which typically differ only at about $O(10^{-8})$).  The individual weights
are seen now to differ quite significantly from the correct value, but the sum of the
three weights does match the correct value, and the contribution of the cluster to
the quadrature is seen to be independent of the size of the cluster.

\section{AN EXPLANATION}

The eigenvalue clustering effect which is well known \cite{ref:parlett}, and the
invariance of weights which we have just described imposes strict constraints
on the mechanism by which roundoff error corrupts the Lanczos procedure.
At finite precision, the Lanczos procedure still generates a sequence of
vectors $f_i, i=1\ldots m$.  However these vectors no longer remain orthogonal,
and no longer tridiagonalise $\mtx{A}$.  We have instead that
\begin{enumerate}
\item
$\mtx{F}_m^\dagger \mtx{F}_m \ne \mtx{I}_m$, 
\item
$\mtx{F}_m^\dagger \mtx{A} \mtx{F}_m = \mtx{H}_m$, symmetric 
but not tridiagonal,
\item
The tridiagonal matrix, $\mtx{T}_m$ which is, in practice,
used to estimate
eigenvalues of $\mtx{A}$ is given by the truncation of 
$\mtx{H}_m$ to tridiagonal form (i.e. the all off-tridiagonal
elements of $\mtx{H}_m$ are set to zero).
\end{enumerate}

Ignoring the very small differences between eigenvalues in a
degenerate cluster, the eigenvalue spectrum of
$\mtx{T}_m$ is given as 
$\{\theta_1(n_1), \ldots, \theta_j(n_j), \theta_{j+1}(1), \ldots \theta_{p}(1)\}$,
where $n_i$ is the multiplicity of the cluster with eigenvalue $\theta_i$, $j$ counts 
the non-unit-multiplicity clusters, and $p$ counts the different distinct
eigenvalues.  We denote the $n_i$ eigenvectors with degenerate eigenvalue $\theta_i$
by $s_i^{(k)}, k=1\ldots n_i$.   The $n$-row by $m$ column matrix $\mtx{F}_m$ relates
these eigenvectors to ``Ritz'' vectors $y_i$ which are $n$-component vectors in the
space spanned by the columns of $\mtx{F}_m$ as $y_i = \mtx{F}_m s_i^{(k)}$.  A well
established result \cite{ref:parlett} 
for the Lanczos procedure is that, in a non-unit-multiplicity
cluster, the eigenvectors $s_i^{(k)}$ all generate the same Ritz vector $y_i$ (approximately).
Further this Ritz vector is a true eigenvector of the matrix $\mtx{A}$, and the corresponding
eigenvalue is a true eigenvalue of $\mtx{A}$ (approximately).  The matrix $\mtx{F}_m$ is
therefore seen to be rank-deficient since there are independent 
linear combinations of columns which equal the same vector. 

We have found \cite{ref:cahill} that the best way to expose this rank
deficiency is to execute a singular value decomposition of
$\mtx{F}_m$.  Using this singular value decomposition, it is possible
to find a special set of orthonormal basis vectors for the space
spanned by the columns of $\mtx{F}_m$.  In this special basis, the
matrix $\mtx{H}_m$ takes a block diagonal form with blocks
corresponding to the non-unit-multiplicity eigenvalues of $\mtx{T}_m$
having entries at all positions equal to the corresponding eigenvalue.
When $\mtx{H}_m$ is truncated to diagonal form in the normal way,
these blocks in the special basis representation become diagonal (the
effect of the truncation is to set all the non-diagonal entries for
the non-unit-multiplicity blocks equal to zero).  We thus obtain the
usual degenerate eigenvalue clusters of the truncated matrix
$\mtx{T}_m$.  The reason that the Lanczos procedure at finite
precision produces good eigenvalue estimates is thus seen to be a
combination of the specially structured nature of the space spanned by
the columns of $\mtx{F}_m$, and of the very special way in which the
truncation to diagonal form acts.  The invariance of the sum of
weights for a cluster is then explained by the observation that this
sum gives the projection of the starting vector $\psi$ seeding the
Lanczos procedure onto the corresponding Ritz vector.


\end{document}